# Domain Structure of a Disoriented Chiral Condensate From a Wavelet Perspective

Zheng Huang[a], Ina Sarcevic[a], Robert Thews[a] and Xin-Nian Wang[b]

[a]*Department of Physics, University of Arizona, Tucson, AZ 85721*

[b]*Nuclear Science Division, MS 70A-3307*

*Lawrence Berkeley National Laboratory, Berkeley, CA 94720*

(October, 1995)

## Abstract

We present a novel method for studying the formation of a disoriented chiral condensate (DCC) in high-energy hadronic and heavy-ion collisions utilizing a discrete wavelet transformation. Due to its salient feature of space-scale locality, the discrete wavelet proves to be very effective in probing physics simultaneously at different locations in phase space and at different scales. We show that the probability distributions of the neutral pion fraction for various rapidity-bin sizes have distinctive shapes in the case of a DCC and exhibit a delay in approaching the Gaussian distribution required by the Central Limit Theorem. We find the wavelet power spectrum for a DCC to exhibit a strong dependence on the scale while an equilibrium system and the standard dynamical models such as HIJING have a flat spectrum.

## I. INTRODUCTION

Recently, it has been suggested that in very high energy hadronic and heavy-ion collisions the rapid expansion of the collision debris in the longitudinal (beam) direction leads to supercooling in the interior of the interaction region, and as a result, domains of the "unconventionally" oriented vacuum configurations allowed by the chiral symmetry may be formed [1,2]. Detection of this interesting phenomenon, the so-called Disoriented Chiral Condensate (DCC) [1,2], would provide valuable information on the vacuum structure of the strong interaction and the nature of the chiral phase transition. Preliminary theoretical investigations on nonequilibrium dynamics using the classical linear $\sigma$-model have found some evidence for the growth of long wavelength pion modes, which may indeed lead to domain structure or cluster formation [4,5]. Although the precise dimension of a typical domain or cluster is still under debate, it seems likely, especially in heavy-ion collisions, that many domains or clusters could be formed in the interaction volume. If there are many uncorrelated small domains, the integrated probability distribution of the neutral pion fraction $f$ emitted from a disoriented region, predicted to be $P(f) = 1/2\sqrt{f}$ [1,2], would become Gaussian. This would follow from the Central Limit Theorem [10] in statistics which states that the probability distribution of the sum of $N$ independent identically distributed variables becomes Gaussian with width $\sim 1/N$ as $N$ increases, independent of the original distribution. This makes the experimental search for the DCC signal a rather difficult task. In order to disentangle the DCC domain structure, we propose a new method which emphasizes not only the behavior of the probability distribution in the full phase space region but also its fluctuation in rapidity $\eta$ or azimuthal angle $\phi$. In other words, one needs to study the "local" properties of the distribution in phase space if the DCC clusters are "localized" objects in phase space.

Normally DCC domains are localized in coordinate space. If they develop collective motion in the course of their time evolution, they should also appear localized in momentum space. The standard "count in cell" technique is to divide the phase space into segments with a given acceptance size $\Delta\eta\Delta\phi$ and to define $f$ *locally* in each cell. Since the location of a DCC cluster in phase space will not be fixed event by event, it is not useful to examine the probability distribution of $f$ in a particular cell. Instead, one may look at the probability of finding a particular value of $f$ in *all* cells of the same size in one event and average over the events. Clearly, the probability distribution will depend on the size of the cell. In addition, the information on the fluctuations with a characteristic correlation scale smaller than the acceptance is entirely suppressed.

In this paper, we propose a novel method to study the DCC domain structure which features both space and scale localities. It is a multiresolution analysis performed by a discrete wavelet transformation (DWT) which has been found effective in systematically detecting structures on various scales in turbulence, astrophysics, and multiparticle productions [11–13]. We demonstrate that the DWT proves to be very useful in identifying and measuring the DCC domain structures *simultaneously* in terms of their size (in scale) and location (in space). Since it is likely that there are other physical scales accompanying the typical DCC domain scale in a physical process, the multiresolution feature of the DWT is essential for identification of the structures of interest. It acts like a mathematical microscope which can zoom in or out to various scales at each location. Due to the completeness and orthogonality of the DWT basis, there will be no information loss.

To demonstrate the proposed technique, we will use the DWT method to study DCC domains in numerical simulations of a classical linear $\sigma$-model in 1+1 dimensions [5], assuming boost invariance of the system. We will analyze the fluctuation of the neutral pion fraction $f$ in the spatial rapidity, $\eta = \frac{1}{2}\ln(t+z)/(t-z)$. The idea demonstrated here can be directly applied to experimental data in momentum phase space once they become available. We will also apply the DWT analysis to the data of the HIJING [9] Monte Carlo simulations, which can be considered as normal "backgrounds" in heavy-ion collisions without DCC formation.

## II. BASICS OF DISCRETE WAVELET TRANSFORMATION

Since more extensive reviews on wavelets exist [15], we summarize only some very basic concepts of the DWT and the idea of a multiresolution analysis by using the well known top-hat function leading to the Haar wavelet. Any one-dimensional sample of a point distribution $f(x)$, such as the rapidity distribution of the neutral pion fraction in the interval $[0,1]$ with resolution $\Delta x$, can be represented as a histogram of $2^J$ bins where $J = \mathrm{mod}(|\ln\Delta x|/\ln 2) + 1$

$$f(x) \equiv f^{(J)}(x) = \sum_{k=0}^{2^J-1} f_{Jk}\phi_{Jk}(x), \tag{1}$$

where $\phi_{Jk}(x)$ are the piecewise constant functions:

$$\phi_{jk}(x) = \begin{cases} 1 & k2^{-j} \le x \le (k+1)2^{-j} \\ 0 & \text{otherwise} \end{cases} . \tag{2}$$

Clearly, $f_{Jk}$ in (1) is the value of $f(x)$ in the $k$-th bin. The family of functions $\phi_{jk}(x)$ can be rewritten as the translation and dilation of a single function $\phi(x)$, called the mother function

$$\phi_{jk}(x) = \phi(2^j x - k), \tag{3}$$

where $\phi(x)$ in this case is the top-hat function: 1 in $[0,1]$ and 0 otherwise.

In $\phi_{jk}(x)$, the index $j$ denotes the resolution scale and $k$ the position at each scale. The idea of multiresolution analysis is to find representations of the sample function $f(x)$ at various scales. Obviously $f_{Jk}$ in (1) is such a representation at the finest scale $J$ and the index $k$ takes the place of the coordinate $x$. In order to find the distribution at the next finest scale $J-1$, one needs to expand $f(x)$ in $\phi_{J-1,k}(x)$. However, it cannot be simply done from (1) since $\phi_{J-1,k}(x)$ is not orthogonal to the function $\phi_{Jk}(x)$ at a finer resolution. To solve the problem, consider a stairwell function $\psi(x)$ defined by

$$\psi(x) = \begin{cases} 1 & 0 \le x < 1/2 \\ -1 & 1/2 \le x < 1 \\ 0 & \text{otherwise} \end{cases} . \tag{4}$$

One can construct a wavelet basis $\psi_{jk}(x)$ by the operation of dilation and translation $\psi_{jk}(x) = \psi(2^j x - k)$. The crucial property is that the functions $\psi_{jk}(x)$ are orthogonal to each other with respect to both the scale index $j$ and the position index $k$. For a given $j$, $\psi_{jk}(x)$'s are also orthogonal to $\phi_{jk}(x)$'s. $\psi_{jk}(x)$ and $\phi_{jk}(x)$ are usually called the wavelet functions and the scaling functions, or the father functions and the mother functions, respectively [14]. It is easy to derive the following recursion relation

$$\phi_{j,2k}(x) = \frac{1}{2}(\phi_{j-1,k}(x) + \psi_{j-1,k}(x)) \quad ; \quad \phi_{j,2k+1}(x) = \frac{1}{2}(\phi_{j-1,k}(x) - \psi_{j-1,k}(x)), \tag{5}$$

indicating that all mother functions $\phi_{jk}(x)$ at scale $j$ can be expressed in terms of the mother and the father functions at scale $j-1$. Using (5), one shows that

$$f^{(j)}(x) = \sum_{k=0}^{2^{j-1}-1} f_{j-1,k}\phi_{j-1,k}(x) + \sum_{k=0}^{2^{j-1}-1} \tilde{f}_{j-1,k}\psi_{j-1,k}(x), \tag{6}$$

where the mother function coefficients (MFC) $f_{j-1,k}$ and the father function coefficients (FFC) $\tilde{f}_{j-1,k}$ are given by

$$f_{j-1,k} = \frac{1}{2}(f_{j,2k} + f_{j,2k+1}) \quad ; \quad \tilde{f}_{j-1,k} = \frac{1}{2}(f_{j,2k} - f_{j,2k+1}). \tag{7}$$

Thus, for the Haar basis, the MFC's are just the average between two bins of the MFC's at the previous scale, while the FFC's are the half difference between two bins. Since $\psi_{jk}(x)$ and $\phi_{jk}(x)$ are orthogonal to each other, the mother function expansion is not mixed with the father function expansion at any scale. Therefore, one can safely define the distribution $f(x)$ in the mother function representation at scale $j \leq J$ by

$$f^{(j)}(x) = \sum_{k=0}^{2^j-1} f_{jk}\phi_{jk}(x), \tag{8}$$

where the MFC's $f_{jk}$'s are calculated from recursion relation (7) starting from the finest scale $J$. One can also define the father function representation of the distribution $f(x)$ at scale $j$

$$\tilde{f}^{(j)}(x) = \sum_{k=0}^{2^j-1} \tilde{f}_{jk}\psi_{jk}(x). \tag{9}$$

From (6), (8) and (9), it is clear that $\tilde{f}^{(j)}(x) = f^{(j+1)}(x) - f^{(j)}(x)$, i.e. the father representation at one scale is just the difference of the mother function representations between the next finer scale and that scale. The FFC's can be calculated using the orthogonality of $\psi_{jk}(x)$ in $j$ and $k$

$$\tilde{f}_{jk} = 2^j \int f(x)\psi_{jk}(x)dx, \tag{10}$$

and the MFC's at various scales can be found from $\tilde{f}_{jk}$ using (6) (note that though (6) and (10) is derived based on the Haar wavelet, they are valid for all discrete wavelets).

The multiresolution analysis is a space-scale decomposition (SSD) where the space corresponds to any physical variable like the rapidity and the scale is the size (Fourier) space. Unlike the Fourier transformation which requires the information in whole physical space, the wavelet SSD requires only the local information in space. Roughly speaking, the multiresolution analysis is a journey down (or up) a hierarchy of scales and to view the "world" in different "eyes" with magnifying or reducing glasses. The mother function representations as the scale becomes coarser can be thought of (roughly) as removing higher frequencies while retaining lower frequencies. The father function representations record the higher frequency fluctuations living on the lower frequency structures at each scale and carry the information on the nature of the fluctuations at each scale.

To analyze a DCC domain structure, we should use a SSD localized in both the physical space and scale space. Although the Haar basis is very intuitive and easy to visualize, the top-hat function is discontinuous and thus not localized in scale (Fourier) space. Important progress was made in the mid 80's to early 90's in finding a continuous discrete wavelet basis that is well localized in Fourier space. Specifically, Daubechies [16] constructed several families of wavelets and scaling functions which are orthogonal, have compact support and are continuous. The father (wavelet) function $\psi(x)$ and the mother (scaling) function $\phi(x)$ are defined as

$$\phi(x) = \sum_m c_m \phi(2x - m) \quad ; \quad \psi(x) = \sum_m (-1)^m c_{1-m} \phi(2x - m), \tag{11}$$

where the coefficients $c_m$'s specify a particular wavelet. The Haar wavelet is a special case when $c_0 = c_1 = 1/2$ and all other $c_m$'s are zero. The simplest wavelet which is localized both in space and in scale is given by $c_0 = (1+\sqrt{3})/4$, $c_1 = (3+\sqrt{3})/4$, $c_2 = (3-\sqrt{3})/4$ and $c_3 = (1-\sqrt{3})/4$. It is often called D4-wavelet.

In our analysis, we shall mainly use the Haar wavelet to obtain the mother function representation owing to its obvious connection to the "count in cell" technique, while use D4-wavelet to investigate the scale dependence of the fluctuations (the father function representation). We also check the reliability of our result by using the D20-wavelet.

## III. DYNAMICAL MODEL OF DCC

### A. The Model

The dynamical process of chiral phase transition is not well understood among many theoretical uncertainties in the description of the transition from the quark-gluon plasma to the hadronic gas. If there is significant supercooling before the plasma hadronizes, the evolution of the chiral order parameter is likely to be out of equilibrium. The linear $\sigma$-model is shown [7] to characterize the basic feature of the chiral phase transition since it is in the same universality class as QCD. It also gives a good description of soft pion dynamics such as pion-pion scattering which dominates the late stages of evolution of the system. In the linear $\sigma$-model, the dynamical field $\Phi$ consists of four components $(\sigma, \pi_1, \pi_2, \pi_3)$ denoting the sigma and pion fields. In high multiplicity events where there are a large number of soft pions produced in high energy collisions, one attempts to describe the system by the classical field theory where the quantum effects are generally suppressed due to a nearly complete occupation of quantum states. The dynamical evolution of the interacting pion system is governed by the classical equation of motion derived from the linear $\sigma$-model. The initial condition can be chosen to represent the possible configurations before the phase transition.

We simulate an ensemble of random fields distributed in space with a geometry consistent with the boost invariance [8]. The field distributions are generated by general Gaussian random distributions with a mean field $\langle\Phi\rangle$ and a variance $\langle\delta\Phi^2\rangle$ which determine whether the system is in or out of equilibrium initially. The mean field which characterizes the order parameter evolves according to the equation of motion in a Hartree approximation

$$\left(\frac{\partial^2}{\partial t^2}-\Delta^2\right)\langle\Phi\rangle=\frac{\partial V_{\rm eff}(\langle\Phi\rangle,\langle\delta\Phi^2\rangle)}{\partial\langle\Phi\rangle}; \tag{12}$$

where the effective potential for the mean field $\langle\Phi\rangle$ $V_{\rm eff}(\langle\Phi\rangle,\langle\delta\Phi^2\rangle)$ depends on the mean field and the fluctuation $\langle\delta\Phi^2\rangle$

$$V_{\rm eff}(\langle\Phi\rangle,\langle\delta\Phi^2\rangle)=\frac{\lambda}{4}(\langle\Phi\rangle^2+\langle\delta\Phi^2\rangle-v^2)^2+H\langle\sigma\rangle, \tag{13}$$

where $\lambda=20$, $v=87.4$ MeV and $H=(119\text{MeV})^3$ [5]. The fluctuation $\langle\delta\Phi^2\rangle$ can be thought of as the "effective temperature" though the temperature is not well defined when the system is out of equilibrium. In the supercooling case, the fluctuation is small (the "effective temperature" is low) so that $V_{\rm eff}$ exhibits the shape of a Mexican hat while the mean field lies on the top of the "hat". As the system evolves, the mean field rolls down from the chirally symmetric false vacuum to the true chirally asymmetric vacuum. For low momentum modes, such a rolling down is not stable. As a result, some domains of correlated pions begin to form due to this instability. This scenario is referred to as the quenching mechanism [3]. An equilibrium evolution is a situation where the initial mean field is given by the equilibrium position of the effective potential at certain "effective temperature". In such a case, there is no initial instability. Whether or not the domain structure can develop in such a case is determined by the subsequent dynamical evolution according to (12).

An important point in the classical simulation is the equal probability of all possible directions in internal isospin space at each space point, i.e., the fields $\pi_1$, $\pi_2$ and $\pi_3$ are distributed with equal probability in the ensemble . We also assume that there is no initial correlation of the pion fields between space locations whose distance is larger than the lattice spacing (which is chosen to be about the pion charge radius $\sim 0.5$ fm). The pion fields, are correlated only within a lattice spacing which represents a minimal correlation length present at all times. The existence of such a correlation in a classical description is unavoidable which may reflect the fact that the pion has a finite size. However, it is not clear whether at such small scale the classical treatment is still valid. We emphasize that strictly speaking our results are meaningful only when the scale involved is much larger than the minimal correlation. For more discussions on the dynamical model, we refer readers to Ref. [4,5].

### B. Signal and Background

We shall compare the probability distribution after some evolution time with the initial random distribution and investigate whether there is an evidence for the significant growth of the correlation. Although our model is far from being realistic in predicting the detailed particle production in high energy collisions, it captures the basic isospin symmetry property of the system. In particular, we shall focus on the ratio of neutral pions to total pions which is defined in the classical model as

$$f = \frac{\pi_1^2}{\pi_1^2 + \pi_2^2 + \pi_3^2}, \tag{14}$$

and assume that the dynamical details of particle production cancel out. The initial random distribution is generated based on the most general consideration of isospin symmetry and statistical principle. It serves as a "background" as it represents a system without any further dynamical evolution. We also use the HIJING [9] Monte Carlo model to generate events and calculate the corresponding neutral pion fraction in different bins of the momentum space, which we take as a standard background to the DCC signals. The HIJING result is found to be compatible with that of our initial random sample and a system evolved form an equilibrium initial condition, since all of them do not have a DCC production mechanism.

## IV. WAVELET VIEW OF DCC CLUSTERS

The central idea of using the wavelet analysis is to attempt an unbiased separation of the DCC physics from the various "backgrounds" which cannot be otherwise easily eliminated. The usual correlation method is not so effective here because the different scale correlations are mixed together in the same phase space. The discrete wavelet transformation (DWT), on the other hand, has an extra dimension – the scale which can be used to separate these correlations. The main source of the "backgrounds" is the fluctuations at small correlation scale due to pion radiation from the uncorrelated regions in space-time, which can coexist with a DCC domain. The background can also come from the partial thermalization of the system leading to a thermal correlation length. Note that the short correlation in isospin space does not correspond to high frequency fluctuation in space-time or large momentum mode, which may otherwise be removed by a momentum cut-off.

Although a two-dimensional wavelet transformation is available, we shall confine ourselves to the one-dimensional case for the purpose of demonstration. We have simulated the evolution of the pion fields using a classical linear $\sigma$-model with an $1+1$ boost invariance [5]. The initial condition is chosen to model a quenching scenario [3] where the system supercools after the QCD phase transition. After time evolving this nonequilibrium state from $\tau_{\rm initial} = 1$ to $\tau_{\rm final} = 5$ fm/c, we form a one-dimensional data array of

$$f(\eta) = \frac{\pi_1^2(\eta)}{\pi_1^2(\eta) + \pi_2^2(\eta) + \pi_3^2(\eta)}, \tag{15}$$

for each event as the spatial rapidity ($\eta$) distribution of the $\pi_1$-fraction $f$, where $\pi_i$ ($i = 1, 2, 3$) are the three Cartesian components of the classical pion field. The lattice spacing in $\eta$ is chosen to be $\Delta\eta = 0.08$. Thus initially the pion fields are automatically correlated within $\Delta\eta$ but completely random at scales beyond $\Delta\eta$. The maximal rapidity interval is chosen to be $[-5, 5]$, with a periodic boundary condition. The total number of bins is $2\eta_{\rm max}/\Delta\eta = 2^7$, so $J = 7$ is the finest scale. The large structure formation from a nonequilibrium evolution is compared to the initial random fluctuations at $\tau_{\rm initial} = 1$ fm/c, and also to data generated by equilibrium evolution to $\tau_{\rm final} = 5$ fm/c.

### A. Mother Function Representation

We display in Figure 1, the mother (scaling) function representation of a typical quenching simulation sample $f(\eta)$ using a Haar wavelet basis. The mother function coefficients (MFC) at different scales are calculated using the existing codes provided in *Numerical Recipes* [17].

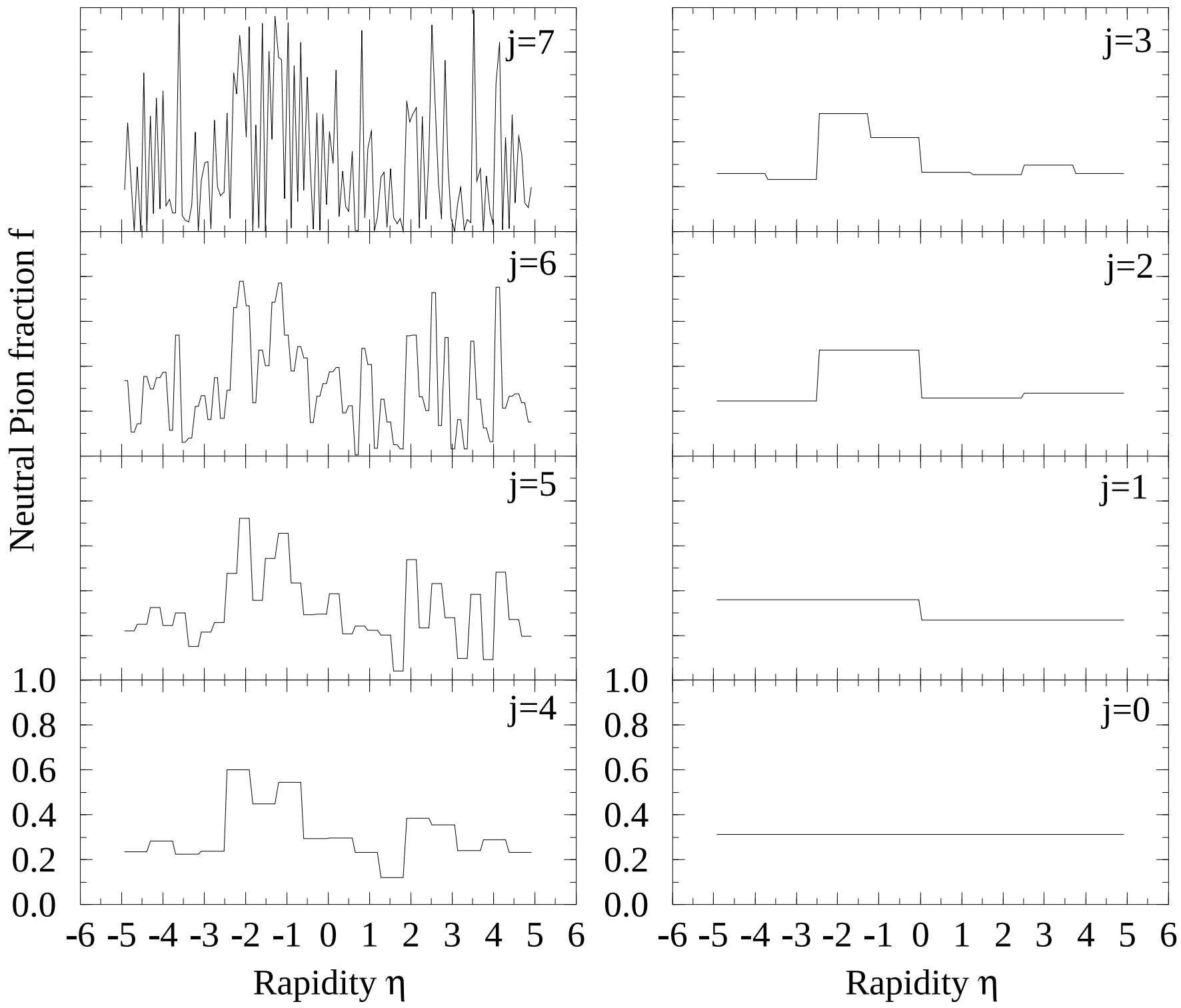

FIG. 1. A mother function representation based on the Haar wavelet multiresolution decomposition. The original sample with a resolution $j = 7$ is the rapidity distribution of the neutral pion fraction $f$ generated by assuming a quenching initial condition. The distributions of the MFC's at $j = 6, 5, ..., 0$ are shown.

At the finest scale $j = J = 7$, the sample contains fluctuations with the correlation sizes ranging from $\Delta\eta$ to $2\eta_{\max}$. Some clustering is evident, most prominently located around $\eta \simeq -2$. High frequency fluctuations superimposed on these cluster domains make it difficult to identify them by eye. At the next finest scale $j = 6$, there are only fluctuations with scales greater than $2\Delta\eta$ present. Any fluctuations with scales smaller than $2\Delta\eta$ have been "filtered" out. In general, at scale $j$, only fluctuations with scales bigger than $2^{J-j}\Delta\eta$ survive. The clustering peaks become more and more distinctive as the scale step $j$ goes down until some scale step $j_d$. For $j$ smaller than $j_d$, the cluster structure cannot survive and will start to disappear. One sees that for this data sample $j_d \sim 3$ to 4. The approximate size of the cluster structure is therefore $\Delta\eta_d = 2^{J-j_d}\Delta\eta \simeq 0.6$ to 1.2 units in rapidity. Interestingly, an even larger structure located at $[-2, 0]$ may be emerging at scale $j = 3$ to 4 and disappears at $j = 1$. At the roughest scale $j = 0$, no structure can exist by definition and only the average value $\langle f \rangle \simeq 1/3$ is plotted.

### B. Delayed Central Limit

The cluster identification shown above is done for a single sample event, which may merely be a statistical fluctuation. Since the locations of the clusters will in general vary from event to event, we cannot reduce the fluctuations by combining many different events. However the probability distribution $P(f)$ should be the same for different events if the DCC domain production mechanism is common to all events. As the second demonstration on the mother function representation, we calculate the probability distribution of the MFC's and average it over 500 simulation events. We study $P^{(j)}(f)$ at different $j$'s, corresponding to the probability distribution at different physical scales, and compare them with the same quantity for a random distribution at the same scale. The random noise samples are generated by using the initial configurations at $\tau_{\text{initial}} = 1$ fm/c in our simulation.

Since our initial fluctuations of $\pi_1$, $\pi_2$ and $\pi_3$ are generated randomly with equal probability for every direction in isospin space, the probability distribution at the finest scale $j = J$ must follow [1,2]

$$\frac{dP^{(J)}}{df}(f) = \frac{1}{2\sqrt{f}} \tag{16}$$

Note that this distribution corresponds to the classical description of the initial state. It does not necessarily mean that the experimental data at finest rapidity resolution should discover such a behavior even for random fluctuations. Our initial correlation length $\Delta\eta \simeq 0.08$ is merely an assumption. In the ideal limit that the detector resolution can be made arbitrarily sharp, at some scale the behavior in (16) may or may not be found depending on whether the number of pions in each bin is large enough so that classical description is still valid. In any case, what remains to be important, as we shall show below, is to discover a behavior similar to (16) well above the finest resolution scale (say, at $\Delta\eta_d \sim 1$ unit).

Assuming that there is no correlation in a noise sample beyond $\Delta\eta$, one can analytically calculate the probability distribution at next scale $j = J-1$ by the convolution of two independent identical distributions at the previous scale $j = J$

$$\frac{dP^{(J-1)}}{df}(f) = 2\int \frac{dP^{(J)}}{df}(f_1)\frac{dP^{(J)}}{df}(2f-f_1)df_1$$
$$= \begin{cases} 2\int_0^{2f} \frac{1}{2\sqrt{f_1}}\frac{1}{2\sqrt{2f-f_1}}df_1 & 0 \le f \le \frac{1}{2} \\ 2\int_{2f-1}^{1} \frac{1}{2\sqrt{f_1}}\frac{1}{2\sqrt{2f-f_1}}df_1 & \frac{1}{2} \le f \le 1 \end{cases} = \begin{cases} \frac{\pi}{2} & 0 \le f \le \frac{1}{2} \\ \sin^{-1}\left(\frac{1-f}{f}\right) & \frac{1}{2} \le f \le 1 \end{cases}, \quad (17)$$

where $f = (f_1 + f_2)/2$. The convolutions become more complicated at subsequent scale steps and we do not attempt to derive them analytically. Nevertheless, we note that the distributions tend to shrink more and more toward the mean value $\langle f\rangle = 1/3$ and their shapes tend to be more and more Gaussian-like as one may expect from the Central Limit Theorem. Indeed, it can be shown [10] that the variance $\langle \Delta f^2\rangle \equiv \langle f^2\rangle - \langle f\rangle^2$ or roughly the width square of the distribution follows a simple rule

$$\langle \Delta f^2\rangle_j = \langle \Delta f^2\rangle_{j-1}/2 , \quad (18)$$

so the distribution tends to be narrower and narrower. We confirm the above picture by calculating the MFC's of the initial random samples and their probability histogram at each scale. The result shown in Figure 2 is averaged over 500 events. The prediction from (17) is also plotted at $j = 6$. The agreement of the random data sample with the prediction indicates that adjacent bins are indeed independent of each other. We have also done a HIJING Monte Carlo calculation for heavy ion collisions at RHIC energy for the rapidity distribution of the neutral pion fraction. The result is very similar to that in Fig.2 for $j \le 5$.

The distribution in (16)at $j = 7$ in Fig.2 results from the minimal correlation of classical fields within the lattice spacing.

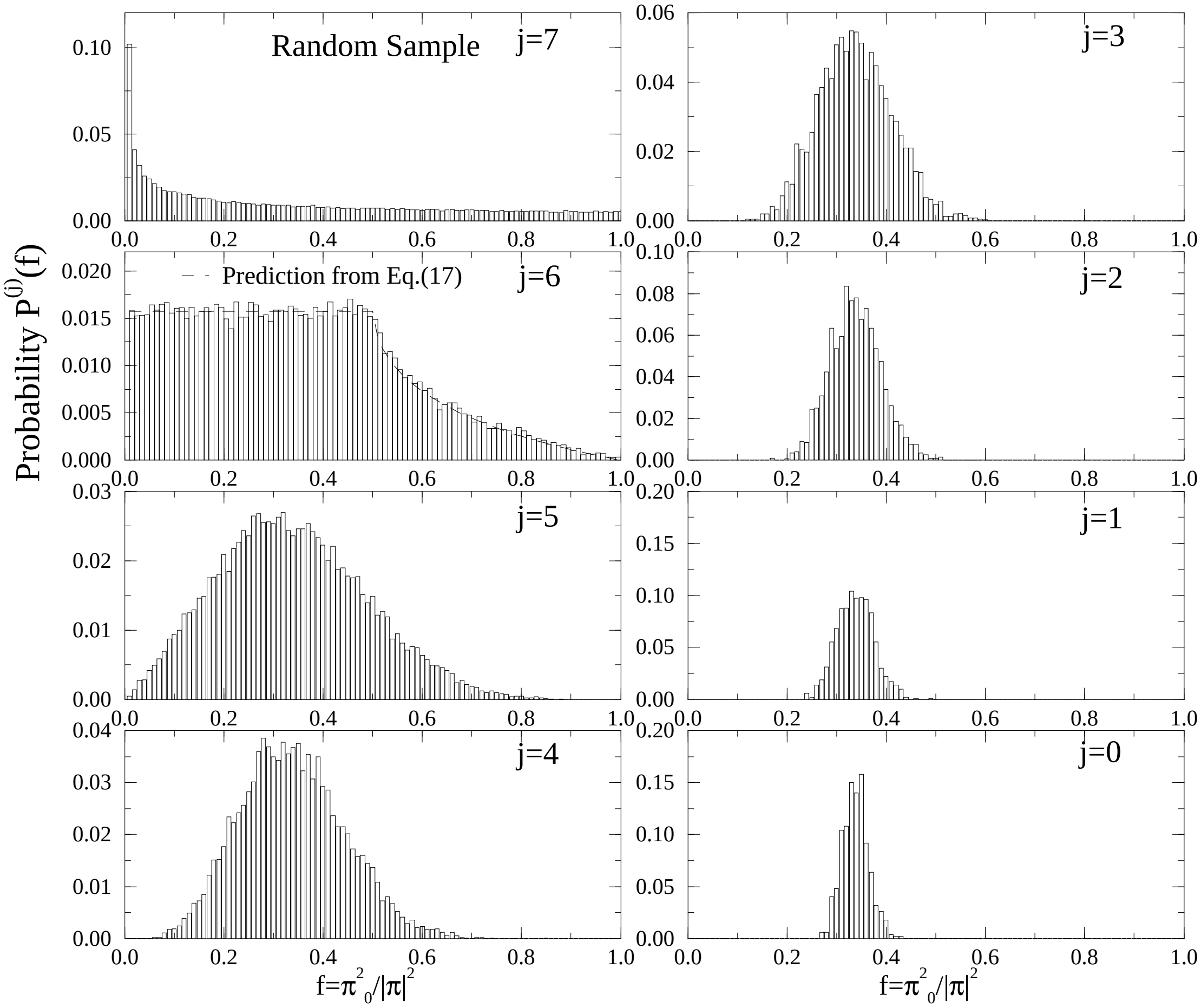

FIG. 2. The probability distribution of the MFC's averaged over 500 random samples at each scale $j$.

The situation for the samples that contain some typical DCC clusters would be very different. In this case, since there exists a correlation scale beyond $\Delta\eta$, the distributions of two adjacent bins are not independent. One expects that the distribution $P^{(j-1)}(f)$ will not be so different from the previous one $P^{(j)}(f)$ until a scale $j_d$. Beyond scale $J_d$, independent convolution is valid. After this the distributions will become sharp according to the rule in (18) provided there are no other interesting scales. The result for 500 quenching samples at $\tau_{\text{final}} = 5$ fm/c is plotted in Figure 3, which shows much broader distributions as compared to the corresponding ones in the random case in Fig. 2.

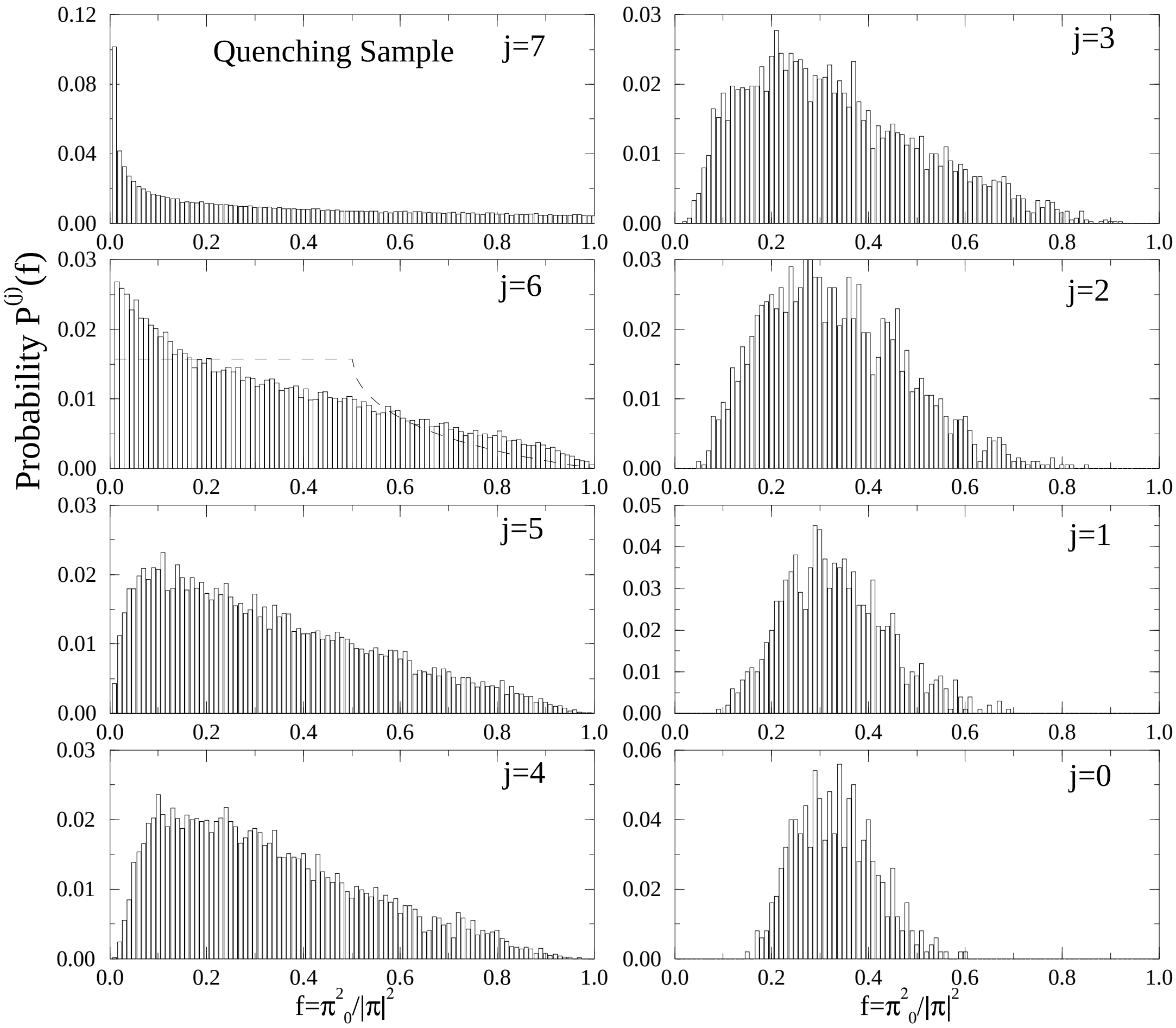

FIG. 3. The probability distribution of the MFC's averaged over 500 DCC samples at each scale $j$.

The distribution given in (17) at $j = 6$ is also plotted (the dashed line) in Fig.3 for a comparison. There is no evidence of flat plateau for $j = 6$ as in Fig.2. This is a good evidence that adjacent bins are very correlated. Only when one reaches $j = 2$ in the DCC case does it show a similar behavior to that at $j = 5$ in the random case. Following this the DCC $j = 1$ is similar to the random $j = 4$, etc., suggesting a phenomenon which we shall refer to as the *Delayed Central Limit*. The "delayed" scales in $j$ are about $\Delta j = 3$ for the probability distribution and are approximately equal to $j_d \simeq 3$ to 4 as inferred from the MFC analysis for a single event shown in Figure 1. This suggests the cause of the delay is due to the DCC clusters. The delay occurs for $j \geq j_d$ and there is no further delay observed after $j_d$ indicating there is no further correlation beyond scale $j_d$. We have also checked simulation samples in the case of an equilibrium evolution where the quenching initial condition is replaced by the annealing initial condition [6] and no significant delay has been found.

### C. Father Function Representation

So far we have discussed the scaling or the mother function representations in the discrete wavelet transformation. The scaling functions, though being very intuitive in the Haar basis, are not orthogonal with respect to $j$. The MFC's at scale $j$ contain information of the MFC's at all subsequent scales. However, the father function coefficients wavelets are orthogonal in both $j$ and $k$ as we have discussed in Sec. II. The information loss in the MFC's as the scale step goes down is recorded in the FFC's at that scale since roughly the FFC's at each scale $j$ are just the difference of the MFC's between this scale and the previous finer scale $j-1$. Therefore, the FFC's at a given scale exclusively describe the nature of fluctuations at that scale and there is no contamination of the information from other scales.

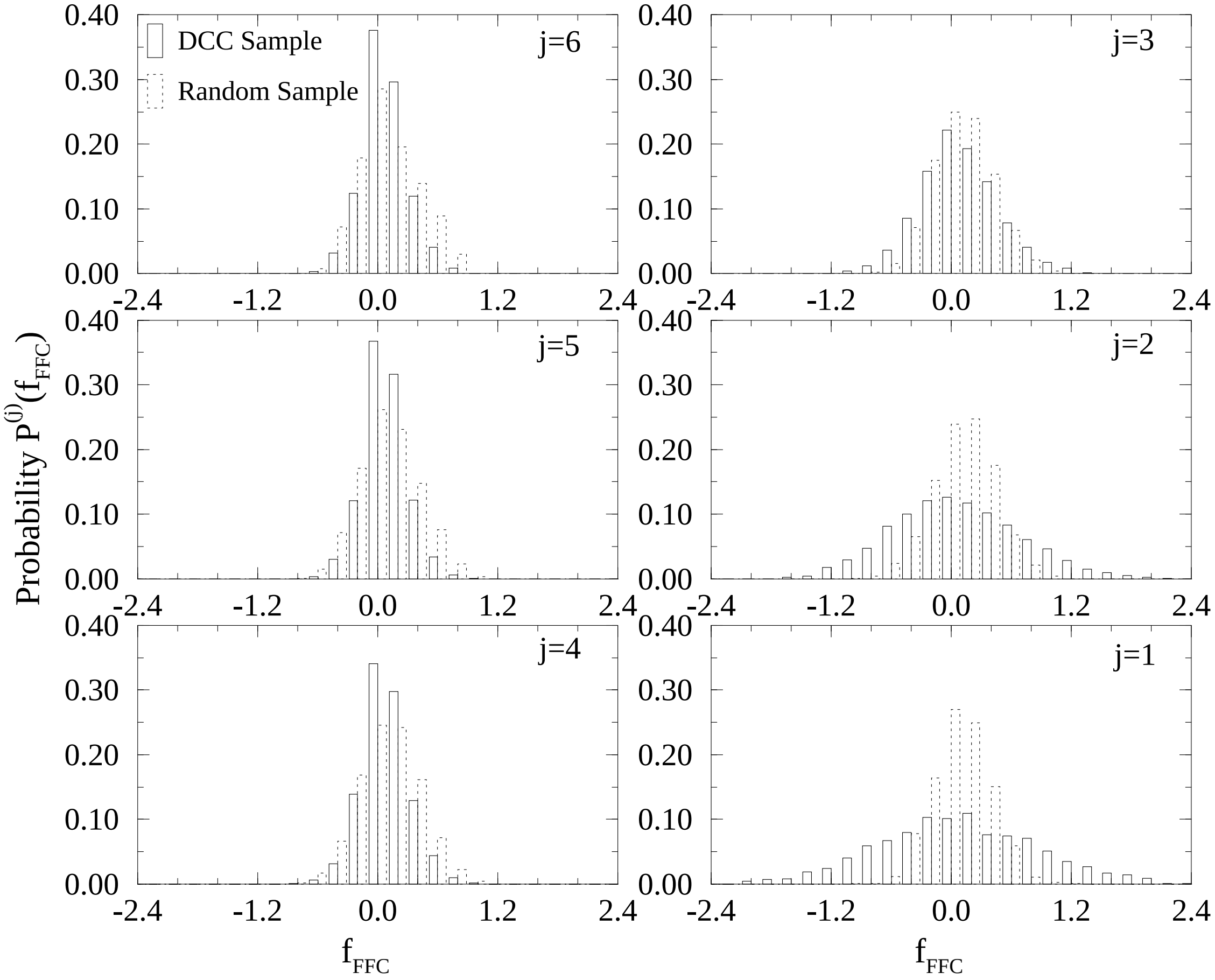

FIG. 4. The probability distributions of the FFC's for both DCC samples and random samples at each scale $j$.

We have calculated the father function representation of 500 DCC samples and 500 random noise samples using D4-wavelet. Again we concentrate on location independent quantities such as the probability distribution of FFC's. At a given scale, unlike the MFC's, the FFC's are not aware of the existence of the clustering structures at the subsequent scales – they only measure the fluctuations with that "frequency" $1/\Delta\eta$. The distributions should not look very different as $j$ decreases toward $j_d$ at which clustering occurs, as the FFC's only record the "noise". They should be quite similar to those of the random noise samples except that the "noise levels" are lower in the DCC samples which implies a smoother rapidity distribution. Thus, they should be more peaked around $\langle \tilde{f} \rangle \simeq 0$. As $j$ decreases further the clusters begin to appear, and fluctuations will grow as the FFC's measure the difference between the structure and the structureless regions. Phrased more intuitively, the FFC's at different scales can be thought of as measuring the fluctuations using various sized "rulers". When the size of a ruler matches the size of a structure, the strength of fluctuations grows and thus the probability distribution of FFC's become more spread-out when $j \leq j_d$. The above picture is supported in our demonstration as shown in Figure 4 where the probability distributions of FFC's at each scale for both the DCC samples and the noise samples are plotted. For the noise samples as one may anticipate, the probability histograms look alike at all scales, indicating the nonexistence of any scale structures. We define $j_d$ to be the scale for the DCC clustering, beyond which the fluctuations surpass the pure random noise fluctuations. Clearly from Figure 4 this happens at a scale between $j = 4$ and $j = 3$.

### D. Power Spectrum of Rapidity Distribution

A more quantitative way to analyze the fluctuations at each scale is the wavelet power spectrum. Similar to the Fourier analysis, a Parseval Theorem for DWT utilizes the complete and orthonormal properties of the DWT:

$$\frac{1}{L}\int_0^L |f(x)|^2 dx = \frac{1}{L}\sum_{j=0}^{\infty}\frac{1}{2^j}\sum_{k=0}^{2^j-1}|\tilde{f}_{jk}|^2 \,, \tag{19}$$

where $\tilde{f}_{jk}$ are the FFC's. Thus, one can define the power spectrum with respect to the wavelet basis at each $j$ as

$$P_j = \frac{1}{2^j}\sum_{k=0}^{2^j-1}|\tilde{f}_{jk}|^2 \,. \tag{20}$$

$P_j$ is then the power of fluctuations on length scale $L/2^j$. It is also the mean square of the probability distributions of FFC's in Fig. 4. The random noise sample features a flat power spectrum, i.e. the power of fluctuations is the same at any scale. For the DCC sample there should be a flat spectrum when $j \geq j_d$ and some power build-up should show up when $j \leq j_d$. We demonstrate the power spectrum for the rapidity distribution of the FFC's for three different dynamical scenarios in Figure 5. Indeed, a plateau structure is evident in Fig. 5 as $j$ goes down from 6 to 4 in the DCC power spectrum. The power level of that plateau is lower than those in the annealing equilibrium evolution scenario and the random noise sample which indicates a smoother distribution within a DCC domain. The power starts to build up as $j$ goes down below 4 and rises rather quickly, departing distinctively from the annealing and the random noise cases. The crossing point in Fig. 5 is found to be at $j_d = 3.6$ which unambiguously suggests the existence of the DCC clustering with a typical size of $\Delta\eta_d = 2\eta_{\max}/2^{j_d} \simeq 0.8$ units in rapidity.

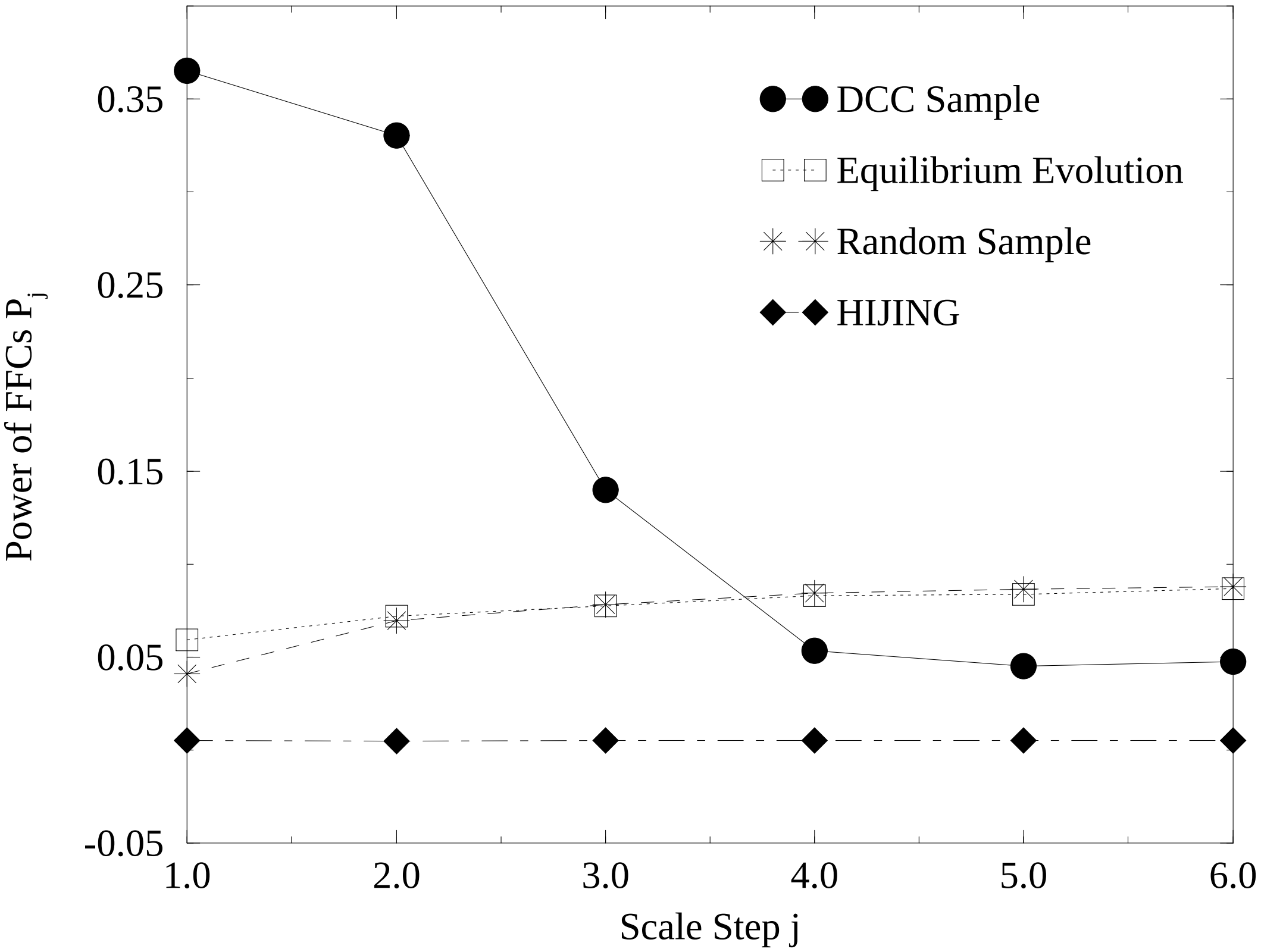

FIG. 5. The power spectra of the FFC's for different dynamical scenarios.

Also plotted is the power spectrum from HIJING Monte Carlo data which is also flat, consistent with the random noise case.

The existence of a plateau structure in the wavelet power spectrum is important in that one may attempt to define an effective "temperature" inside a domain structure where the fluctuations are relatively "stable" against the scale change. The power spectrum can be thought of describing the scale dependence of the "temperature". The annealing scenario (following an equilibrium evolution) or the noise sample features a uniformly flat spectrum and thus possesses a universal (global) temperature; while the rapid rise of the spectrum in the DCC case reflects the highly nonequilibrium nature of the fluctuations and the sensitivity of the temperature on the scale change.

## V. CONCLUSIONS

We have shown that the discrete wavelet transformation method can be an efficient and reliable tool for identifying and measuring the clustering structures characteristic to the formation of the disoriented chiral condensate in high energy hadronic and heavy-ion collisions. The DWT provides a vehicle for discovering physics at different scales that would otherwise be hard to detect, and opens a window to study the scale-dependence of various features of clustering in the DCC. We have demonstrated that a DCC with a typical cluster size of 1 unit in rapidity can be unambiguously identified using both the mother and the father function representations. The DWT method can be very effective in detecting the nature of the fluctuations at different scales as to whether they follow an equilibrium or a nonequilibrium distribution. Furthermore, one can use the DWT to analyze the real space-time structure of the classical field evolution. This will be studied further in a separate publication.

## ACKNOWLEDGMENTS

We are in debt to Peter Carruthers, Li-Zhi Fang and Jesus Pando for their introduction of the wavelet concept to us and many discussions on the technical details. Z.H and X.-N.W thank M. Asakawa for useful discussions. This work was supported in part through the U.S. Department of Energy under Contracts Nos. DE-AC03-76SF00098, DE-FG03-93ER40792 and DE-FG02-85ER40213.